# Fast all-optical random number generator

JIAMIN ZHAO,[1] PU LI,[1,2,3,4,*] XING ZHANG,[5] ZHENSEN GAO,[2,3] ZHIWEI JIA,[1] ADONIS BOGRIS,[6] K. ALAN SHORE,[7] AND YUNCAI WANG[2,3]

[1]Key Laboratory of Advanced Transducers and Intelligent Control System, Ministry of Education, Taiyuan University of Technology, Taiyuan 030024, China
[2]School of Information Engineering, Guangdong University of Technology, Guangzhou 510006, China
[3]Guangdong Provincial Key Laboratory of Photonics Information Technology, Guangzhou 510006, China
[4]Key Laboratory of Specialty Fiber Optics and Optical Access Networks, Shanghai University, Shanghai 200444, China
[5]Changchun Institute of Optics, Fine Mechanics and Physics, Chinese Academy of Sciences, Changchun 130033, China
[6]Department of Informatics and Computer Engineering, University of West Attica, Athens 12243, Greece
[7]School of Electronic Engineering, Bangor University, Wales LL57 1UT, United Kingdom
*Corresponding author: lipu8603@126.com

**We propose a simple and all-optical method for fast random number generation based on the laser mode hopping. Through periodically restarting a two-mode laser operating in the bistable state, a random number stream can be generated due to the spontaneous emission noise. To validate the feasibility of this method, we perform a theoretical simulation using the common vertical-cavity surface-emitting laser (VCSEL) with two polarization modes. Numerical results demonstrate that fast 2.5 Gb/s random number streams can be continuously obtained with verified randomness. This scheme provide a fully monolithic solution for random number generator, due to its simple and all-optical structure.**

Random number generators (RNGs) play crucial roles in varies of fields ranging from scientific calculation to information technology. Especially, the problem of unconditionally secure communications has led to the proposal of high-speed Gb/s physical random number generator in recent years [1-15].

Utilizing stochastic dynamics in lasers to generate physical random numbers has been viewed as a promising solution for this issue [2-15], since Uchida *et al.* firstly reported 1.7 Gb/s physical random number generation using two chaotic laser diodes in 2008 [1]. For instance, Li *et al.* demonstrated real-time ultrafast photonic random number generation using a bandwidth-enhanced chaotic laser [7]. Williams *et al.* and Li *et al.* reported Gb/s physical random number generation by means of filtered amplified spontaneous emission (ASE) from a fiber amplifier [9] or a superluminescent LED [10], respectively. Qi *et al.* and Guo *et al.* demonstrated that high-speed random numbers can be generated by measuring phase noise of a distributed-feedback (DFB) diode laser [11] or a vertical-cavity surface-emitting laser (VCSEL) [12], respectively. Gabriel *et al.* proposed to generate unique random numbers based on quantum vacuum states [13]. Nie *et al.* presented practical and fast random number generation based on photon arrival time [14].

However, most of reported photonic schemes for physical random number generation require bulky and discrete photonic and electronic devices (laser diodes, photodetectors, ADC, logic gates, FPGA, etc.). Generally, these schemes contain four parts: laser sources, sampling, quantizing and post-processing modules. As a result, these systems only demonstrate the feasibility of the functionalities rather than an expecting prospect of compact form and monolithic integration. Moreover, the power consumption of the system increases along with increases in cost in order to support so many components.

On the other hand, mode hopping is another common stochastic phenomenon in different kinds of lasers including vertical-cavity surface-emitting lasers (VCSELs) [16], semiconductor ring lasers [17], $CO_2$ lasers [18], fiber lasers [19] and two-dimensional micro-lasers [20]. Especially, the instabilities between the two orthogonal polarization modes in VCSELs attracted much attention. For instance, Olejniczak *et al.* analyzed polarization switching and polarization mode hopping in VCSELs [21]. Sciamanna *et al.* studied the random mode hopping in VCSELs subject to optical feedback [22]. Coarer *et al.* reported noise-induced attractor hopping in VCSELs under optical injection [23]. Virte *et al.* demonstrated the polarization chaos generation from a free-running VCSEL [24].

Herein, we present a proof-of-principle demonstration for direct generation of high-speed physical random number streams using mode hopping in a solitary two-mode laser. Specifically, we periodically switch the operation of VCSEL from an un-lasing regime to a mode-hopping regime by modulating its injection current with a square wave. After balancing the asymmetry between two orthogonal polarization modes utilizing spontaneous emission noise, the lasing state of VCSEL will unpredictably settle in one of the two polarization modes during each modulation period. Consequently, this stochastic mode selection cause the final random number sequence.

Our simulation results demonstrate that at least 2.5 Gb/s random number stream can be directly generated using our method. The resulting random number stream resembles a perfect coin toss and passes all the standard benchmark tests. In comparison with the aforementioned RNGs [2-15], our RNG system is constructed using just one component: a solitary VCSEL laser diode operating in a bistable regime. In consequence our RNG has a simple and compact form with the potential to be monolithic integrated. Moreover, the present RNG has 'all-optical' functionality omitting any photo-electronic conversion. This means that the final binary random bit streams can be directly obtained at the output port of the VCSEL laser diodes in real time. Thus, this RNG is especially useful for future generation all-optical communication systems.

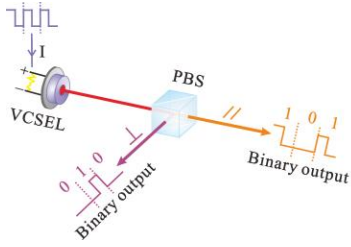

**Fig. 1.** Schematic of random number generation based on a solitary VCSEL. VCSEL, vertical-cavity surface-emitting laser; PBS, polarizing beam splitter; //, parallel LP mode (X-LP); ⊥, vertical LP mode (Y-LP).

Figure 1 depicts the schematic for random number generation based on a solitary VCSEL. The output light of the VCSEL is divided into two orthogonal linear polarization (LP) modes by a polarizing beam splitter (PBS): one is the parallel LP mode (X-LP) and the other is the vertical LP mode (Y-LP). At the same time, the injection current of the VCSEL is modulated by a square wave signal to periodically restart the lasing state. Due to the existence of spontaneous emission noise, the VCSEL will randomly operate at either the X-LP or Y-LP mode during each lasing period. Finally we can get two streams of complementary binary random bits at the two output of the PBS, respectively.

Specifically, we use the spin-flip model (SFM) with noise terms to analyze the behavior of the solitary VCSEL as follows [25, 26].

$$\frac{dE_x}{dt} = \kappa(1+i\alpha)(NE_x - E_x + inE_y) - (\gamma_\alpha + i\gamma_p)E_x + F_x \quad (1)$$

$$\frac{dE_y}{dt} = \kappa(1+i\alpha)(NE_y - E_y - inE_x) + (\gamma_\alpha + i\gamma_p)E_y + F_y \quad (2)$$

$$\frac{dN}{dt} = -\gamma_N N(1+|E_x|^2+|E_y|^2) + \gamma_N \mu - i\gamma_N n(E_y E_x^* - E_x E_y^*) \quad (3)$$

$$\frac{dn}{dt} = -\gamma_s n - \gamma_N n(|E_x|^2+|E_y|^2) - i\gamma_N N(E_y E_x^* - E_x E_y^*) \quad (4)$$

$$F_x = \sqrt{\beta_x \gamma_N /2}\left(\sqrt{N+n}\xi_+ + \sqrt{N-n}\xi_-\right) \quad (5)$$

$$F_y = i\sqrt{\beta_y \gamma_N /2}\left(\sqrt{N-n}\xi_- - \sqrt{N+n}\xi_+\right) \quad (6)$$

Herein, $E_x$ and $E_y$ are two orthogonal linearly polarized field amplitudes, respectively. $N$ represents the total carrier inversions between conduction and valence bands, whilst $n$ is the difference between carrier inversions with opposite spins. Note that $F_x$ and $F_y$, the noise terms arising from spontaneous emission, have been normalized by an integration step $\Delta t$ =10 ps as in Refs. [26-29]. This normalization is to ensure the spontaneous emission amplitudes are not dependent upon the integration time-step in the simulation.

$\beta_x$ and $\beta_y$ are the fraction of the spontaneously emitted photons that goes into the X-LP and Y-LP modes, respectively. $\xi_\pm$ are the Gaussian white noise sources with zero mean and correlation $\langle\xi_\pm(t)\xi^*_\pm(t')\rangle = 2\delta(t-t')$. Other parameters and their values used in simulation are field decay rate $\kappa$=300 ns$^{-1}$, linewidth enhancement factor $\alpha$=3, carrier decay rate $\gamma_N$=10 ns$^{-1}$, spin-flip relaxation rate $\gamma_S$=1000 ns$^{-1}$, linear birefringence $\gamma_p$ =60 ns$^{-1}$, linear dichroism $\gamma_\alpha$ =-0.1 ns$^{-1}$, normalized injection current $\mu$ ($\mu$=1 at threshold), and central wavelength $\lambda$=850 nm. These parameters are selected based on the actual VCSEL devices in recent development. Especially, $\gamma_N$ is fixed according to a recent experimental report [30].

To characterize the different operation states of the VCSEL, we firstly analyze the stability diagram for the two linearly polarized solutions in the absence of noise terms. As shown in Fig. 2, the parameter space ($\gamma_p$, $\mu$) is divided into four regions by the two stability boundaries [i.e., $\mu_x$ and $\mu_y$ given in Eq. (7)] for the X-LP and Y-LP solutions, respectively. When the injection current $\mu$ increases above the boundary line $\mu_x$, the X-LP mode become unstable. On the contrary, the un-stability of the Y-LP mode occurs when the injection current $\mu$ decreases below the boundary $\mu_y$. Specifically speaking, the Y-LP mode is stable in the blue region; in the green region, the X-LP is stable; in the pink region, no LP mode is stable; in the yellow region, both LP modes are stable. In addition, the VCSEL operates in the un-lasing state in the gray region because the injection current is below than the lasing threshold.

$$\mu_x \approx 1+\frac{\gamma_s(\gamma_p^2+\gamma_\alpha^2)}{\gamma_N\left[\kappa(\gamma_\alpha+\alpha\gamma_p)-\gamma_p^2\right]}, \mu_y \approx 1-\frac{2\gamma_\alpha(\gamma_s^2+4\gamma_p^2)}{\kappa\gamma_N(\gamma_s-2\alpha\gamma_p)} \quad (7)$$

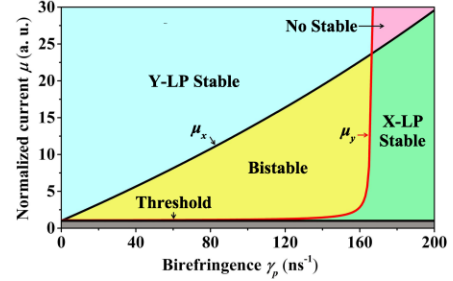

**Fig. 2.** Stability diagram of the X-LP and Y-LP of the solitary VCSEL in the parameter space ($\gamma_p$, $\mu$): in the blue region, the Y-LP mode is stable; in the green region, the X-LP is stable; in the pink region, no LP mode is stable; in the yellow region, both LP modes are stable.

Furthermore, our attention focuses on the bistable region, because only in this region the VCSEL occurs mode-hopping used for random number generation. Figure 3 illustrates the L-I characteristics of the VCSEL output, where the black and red solid lines correspond with X-LP and Y-LP outputs, respectively. At the beginning, when the injection current is relative small (1<$\mu$<1.2), the X-LP mode first outputs and remains stable no matter whether the spontaneous emission noise is considered [Fig. 3(a)] or not [Figs. 3(b) and 3(c)]. With the increase of injection current, the VCSEL will then enter into the bistable region (1.2<$\mu$<8.1). In the absence of noise, a fixed polarization mode oscillates at each injection current, as shown in Fig. 3(a). However, when the spontaneous emission noise is considered, the VCSEL will be initiated to mode hopping between the two LP modes. This can be confirmed by contrast the two L-I characteristics in Fig. 3(b) and Fig. 3(c), which are obtained from the same initial conditions and the same noise strength. As seen, the mode hopping is characterized by the randomly complete exchange of intensity between the X-LP and Y-LP modes. For instance, at the injection current $\mu = 6$, the X-LP mode is dominant in Fig. 3(b), but the Y-LP become the dominant in Fig. 3(c). This implies that a random number stream can be directly obtained from the VCSEL output if the laser is repeatedly reset to such an unstable point in the bistable regime.

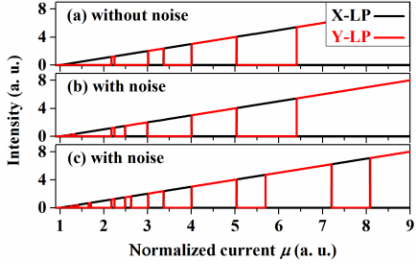

**Fig. 3.** L-I characteristic curves of the VCSEL. (a) L-I curve of the VCSEL without noise; (b)-(c) Two L-I curves of the VCSEL starting from the same initial conditions and noise strength.

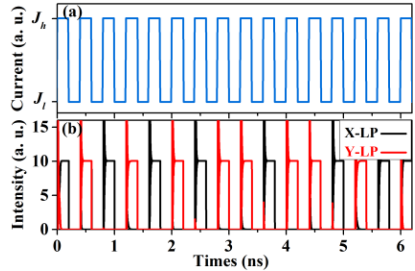

**Fig. 4.** The result of the generated 2.5 Gb/s random pulse sequence. (a) 2.5 GHz square wave current modulation signal, $J_l = 0.98$ and $J_h = 6$; (b) Time series of the X-LP (black) and Y-LP (red) light intensities.

In the simulation, we use a square wave current to periodically modulate the VCSEL to demonstrate the random number generation. Figure 4(a) shows the temporal waveform of the used square wave signal with a repetition rate of 2.5 GHz. Herein, the low current level $J_l = 0.98$ is set below the threshold current $\mu = 1$, while the high level $J_h = 6$ is just the aforementioned unstable point separating the basins of the two LP states. As shown in Fig. 4(b), the final LP state of the VCSEL output will randomly locate at one of the X-LP and Y-LP modes in each modulation period. Note, the black and red waveforms are the associated X-LP output and the Y-LP output, respectively. Herein, the random pulse outputs are just the random number streams: when there is pulse output, it is coded as logical "1" output; otherwise it is coded as logical "0" output. Moreover, it should be mentioned that the final generation rate of random numbers is 2.5 Gb/s, which is controlled by the repetition rate of modulating square wave signal.

Figure 5(a) shows the autocorrelation function (ACF) of the generated random number stream. The AC coefficient level near zero means that the correlation of the generated random number is statistically insignificant. Figure 5(b) depicts the probability density function (PDF) of the random number sequence with an exactly symmetrical distribution, which indicates a statistically unbias. Both the ACF and PDF roughly enables the high randomness of the generated random numbers.

To qualify the randomness of the random bits more stringent, we use the state-of-the-art NIST test suite with 15 statistical tests [31] that is proposed to determine whether a RNG is suitable for cryptographic applications. Figure 5(c) shows the NIST test results. As advised by the NIST, each test is performed using 1000×1 Mbits with a statistical significance level $\alpha=0.01$. For success, the proportion of the sequences satisfying condition should be in the range of 0.99±0.0094392, and the uniformity P-value should be larger than 0.0001. It can be confirmed from Fig. 5(c) that the random number sequences pass all of the NIST tests successfully.

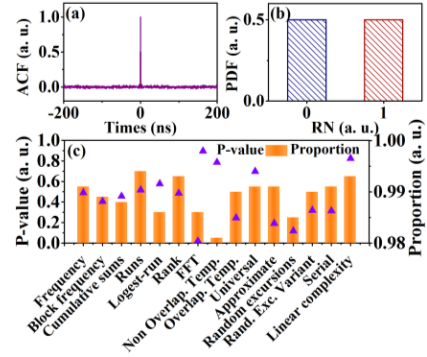

**Fig. 5.** Randomness of the obtained random number (RN) stream: (a) Autocorrelation function (ACF); (b) Probability density function (PDF); (c) NIST test results: P-value (left column) and Proportion (right column). Note, the 15 test items are shown along the horizontal axis.

Herein, we want to point that the spontaneous emission noise is critical to ensure the uniformity of the generated random numbers. The actual VCSEL device has a preferred output mode due to the material non-uniformity. In the model, the dichroism $\gamma_a$ and the birefringence $\gamma_p$ describe the associated amplitude and phase anisotropy induced by the material anisotropy, respectively. Their values in our simulation are $\gamma_a=-0.1$ns$^{-1}$ and $\gamma_p=60$ns$^{-1}$. In this situation, the X-LP mode will be the priority output mode. When the influence of spontaneous emission noise is small [i.e., the value of $\beta_{x,y}$ is small], the laser maintains the output X-LP mode. Until $\beta_{x,y}=2\times10^{-4}$, the mode hopping occurs in the bistable region. The probability of the appearance of the X-LP and Y-LP modes are calculated to be 59.8 % and 40.2 %. Note that the minimum noise strength $\beta_{x,y}$ to achieve mode-hopping is strongly influenced by $\gamma_a$ and $\gamma_p$, as shown in Figs. 6 (a) and 6(b). Through decreasing $\gamma_a$ and $\gamma_p$, $\beta_{x,y}$ can be reduced from the order of 10$^{-4}$ to 10$^{-6}$. For unbiased random number generation, the amount of the spontaneous emission noise coupled to the Y-LP mode should be enhanced to reduce the asymmetry of the output ratio of the two modes. As shown in Fig. 6(c), the frequency of the Y-LP mode monotonically increases with the decrease of signal-to-noise ratio (SNR) [i.e., the enhancement of the noise intensity]. When the SNR is located around 1.72 dB, the frequency of the appearance of the Y-LP mode is close to 50 %.

In addition, it should be noticed that the generation rate of 2.5 Gb/s in our principle-of-proof demonstration is not the ultimate speed. In our scheme, different rates of random numbers can be obtained by controlling the modulation frequency of the square-wave signal. But, the modulation period cannot be decreased indefinitely. The ultimate rate is limited by the switching time of the polarization mode [*i.e.*, the transition time required by the laser to switch from a non-lasing mode to a stable lasing mode]. This can be confirmed from the enlarged waveform of random number stream. From Figs. 6(d)-(e), one can clearly observe that every time the VCSEL is abruptly biased from $J_l$=0.98 to $J_h$=6 at t=0, the final lasing mode has to experience a period of relaxation oscillation to reach a stable output. Therefore, the half-period of the modulating signal needs to be larger than the transition time to obtain a distinguishable bit waveform. In fact, the relaxation oscillation of VCSEL can be further improved by optimizing the internal parameters. According to the expression of relaxation oscillation $f = \sqrt{(2\kappa\gamma_N(\mu - 1))}/2\pi$, one can appreciate that with increase of the normalized injection current $\mu$, the field decay rate $\kappa$ and the carrier decay rate $\gamma_N$, the relaxation oscillation can be effectively increased. The modulation bandwidth of available VCSELs has reached 35.5 GHz [32], so it can expect that using our method it is possible to reach a generation rate at the level of tens of Gb/s through optimizing the internal structure of the laser chip.

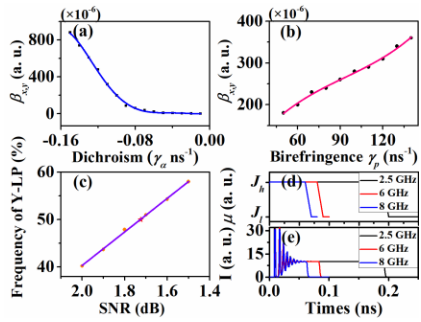

**Fig. 6.** (a)-(b) The minimum noise strength $\beta_{xy}$ to force mode-hopping versus the dichroism $\gamma_\alpha$ and birefringence $\gamma_p$; (c) The frequency of the Y-LP mode versus the noise intensity coupled to the Y-LP mode; (d) Modulating square-wave signals with a frequency of 2.5 GHz, 6 GHz, and 8 GHz; (e) Random number waveforms with a generation rate of 2.5 Gb/s, 6 Gb/s, and 8 Gb/s.

In conclusion, we have proposed theoretically a simple method to directly generate physical random numbers using the mode-hopping in two-mode lasers. Numerical results demonstrated that at least 2.5 Gb/s fast random number streams can be continuously generated by periodically resetting the VCSEL with two LP modes. This scheme is simply implemented by only a solitary laser and thus greatly reduce the system complexity of current RNGs including sampling, quantizing and post-processing modules. Moreover, the proposed RNG has 'all-optical' functionality omitting any particular photo-electronic conversion. This means that our work provides a new way for random number generation, which is suitable for monolithic integration and directly compatible with future all-optical communication.

**Funding.** National Natural Science Foundation of China (NSFC) (61775158, 61961136002, 61927811, U19A2076, 61705159, 61805168, 17174343, 11904157); National Cryptography Development Fund (MMJJ20170127); China Postdoctoral Science Foundation (2018M630283, 2019T120197); Natural Science Foundation of Shanxi Province (201901D211116); STCSM (Grant No. SKLSFO2018-03); Project of Key Laboratory of Radar Imaging and Microwave Photonics (Nanjing University of Aeronautics and Astronautics), Ministry of Education (No. RIMP2019002); Program for the Top Young Academic Leaders of High Learning Institutions of Shanxi; Program for Guangdong Introducing Innovative and Enterpreneurial Teams.

**Disclosures**. The authors declare no conflicts of interest.